\documentstyle[bo99,epsfig]{article}

\title{X-ray Absorption in Radio-Quiet QSOs}
\author{W.N. Brandt,$^1$ S.C. Gallagher,$^1$ A. Laor$^2$ and Beverley J. Wills$^3$}
\affil{1) Department of Astronomy \& Astrophysics, 525 Davey Lab, 
The Pennsylvania State University, University Park, Pennsylvania 16802 USA\\
2) Physics Department, Technion, Haifa 32000 Israel\\
3) Department of Astronomy, University of Texas at Austin,
Austin, Texas 78712 USA}

\def\asca{{\it ASCA\/}}
\def\bepposax{{\it BeppoSAX\/}}
\def\chandra{{\it Chandra\/}}
\def\einstein{{\it Einstein\/}}
\def\rosat{{\it ROSAT\/}}
\def\xmm{{\it XMM\/}}

\def\ltsima{$\; \buildrel < \over \sim \;$}
\def\simlt{\lower.5ex\hbox{\ltsima}}
\def\gtsima{$\; \buildrel > \over \sim \;$}
\def\simgt{\lower.5ex\hbox{\gtsima}}

\def\aox{{$\alpha_{\rm ox}$}}

\begin{document}

\maketitle

\begin{abstract}
Major flows of ionized gas are thought to be present in the nuclei of most 
luminous QSOs, and absorption by these flows 
often has a significant effect upon the observed
X-ray continuum from the black hole region. We briefly discuss X-ray studies of 
this gas and attempts to determine its geometry, dynamics, and ionization physics. 
Our focus is on X-ray warm absorber QSOs, Broad Absorption Line QSOs, and
soft X-ray weak QSOs. We also discuss some prospects for further study with the 
next generation of X-ray observatories. 

\end{abstract}

\section{Introduction}

X-ray absorption studies of active galaxies are proving to be one of the 
most powerful ways of probing material in the immediate vicinity 
of supermassive black holes. Rapid X-ray variability suggests that
the nuclear X-ray source is the most compact 
emitter of continuum radiation, and it thus provides 
a point-like and luminous `flashlight' right at the heart of the 
active galaxy. Because X-rays are highly penetrating, X-ray spectra can 
be used to probe column densities over the wide range 
$10^{19}$--$10^{25}$~cm$^{-2}$. X-ray absorption is
produced by the innermost electrons of metals, and it provides a 
probe of matter in nearly all forms (i.e., neutral gas, ionized gas, 
molecular gas and dust). 

Here we shall briefly review several types of X-ray 
absorption seen in luminous, radio-quiet QSOs. We will discuss
X-ray warm absorber QSOs, Broad Absorption Line QSOs, and
soft X-ray weak QSOs. We will also discuss some future prospects for 
radio-quiet QSO X-ray absorption studies. Due to lack of space, we 
will not be able to cover the important red QSO and type~2 QSO debates 
or the exciting recent results on X-ray absorption in radio-loud QSOs. 

\section{X-ray Warm Absorbers in Radio-Quiet QSOs}

Warm absorption by ionized nuclear gas is familiar from the lower-luminosity 
Seyfert~1 galaxies where it has been intensively studied (e.g., Reynolds 1997; 
George et~al. 1998). Warm absorbers imprint moderately strong edges 
(e.g., O~{\sc vii} and O~{\sc viii}) on the continuum, 
but this absorption is not so strong
that it completely extinguishes the soft X-ray flux. Assuming
photoionization equilibrium, the column density and ionization parameter 
of the ionized gas can be obtained via X-ray spectral fitting. 

\begin{figure}[t!]
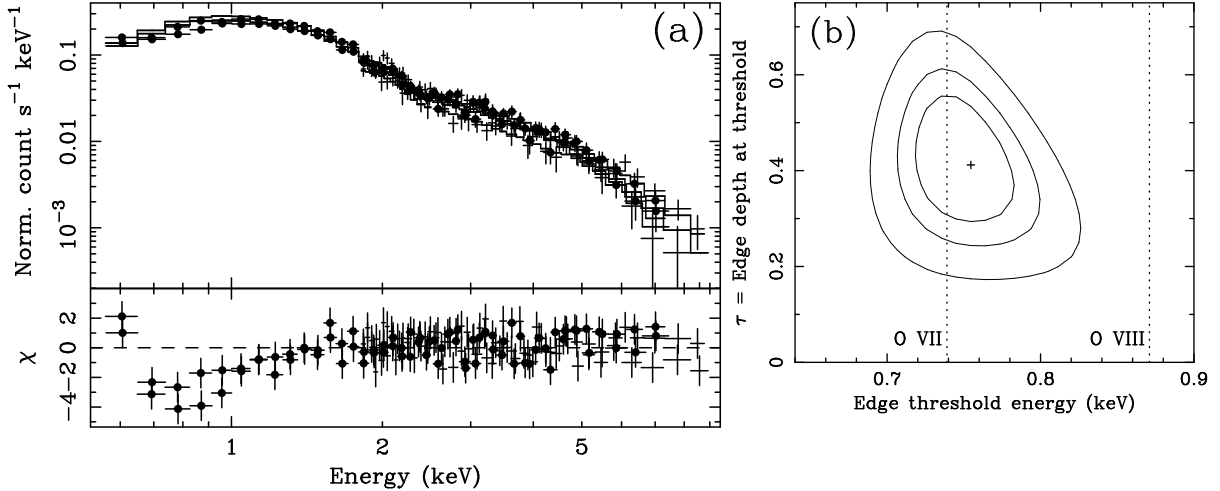

\centerline{\psfig{file=fig1a.ps, width=6.5cm, angle=-90}
\psfig{file=fig1b.ps, width=5.5cm, angle=-90}}
\caption[]{(a) \asca\ SIS (solid dots) and GIS (plain crosses)
observed-frame spectra of the warm absorber QSO IRAS~13349+2438
($z=0.107$). A power-law model has been fit to the 2--9~keV data 
and extrapolated back to show the deviations at 
low energies. The ordinate for the lower panel (labeled $\chi$) shows the 
fit residuals in terms of $\sigma$ with error bars of size one. Note the 
systematic absorption residuals at low energies
due to ionized oxygen. 
(b) Confidence contours for the edge parameters. Contour levels
are for 68.3, 90.0 and 99.0\% confidence. The fitted edge energy has
been corrected for cosmological redshift. 
From Brandt et~al. (1997).}
\end{figure}

To our knowledge, only five luminous radio-quiet QSOs have been shown
to have X-ray warm absorbers: 
MR~2251--178 (e.g., Halpern 1984; Pan, Stewart \& Pounds 1990; Reeves et al. 1997), 
IRAS~13349+2438 (Brandt, Fabian \& Pounds 1996; Brandt et~al. 1997; 
Siebert, Komossa \& Brinkmann 1999; see Figure~1),  
PG~1114+445 (Laor et~al. 1994; George et~al. 1997), 
IRAS~17020+4544 (e.g., Leighly et~al. 1997), and 
IRAS~12397+3333 (e.g., Grupe et~al. 1998). 
It has been difficult to perform detailed studies of the warm absorbers in these 
QSOs due to limited photon statistics, but the basic physical properties of 
their warm absorbers appear similar to those seen in Seyfert~1s. 
Edges from O~{\sc vii} and O~{\sc viii} seem to be the strongest 
spectral features, and column densities of $\approx 10^{21}$--$10^{23}$~cm$^{-2}$ 
and ionization parameters of $\xi\approx$~20--160~erg~cm~s$^{-1}$ are inferred. 
In three cases (IRAS~12397+3333, IRAS~13349+2438 and IRAS~17020+4544), the 
warm absorber probably contains dust which causes significant reddening of the 
optical continuum. Dust will not be rapidly sputtered at warm absorber 
temperatures (the gas temperature is $\sim 10^5$~K for a photoionized warm 
absorber), and it will not be sublimated if the warm absorber
is located outside the Broad Line Region (BLR). Two 
of the QSOs with X-ray warm absorbers (as well 
as most of the Seyfert~1 galaxies; Crenshaw et~al. 1999) show UV absorption 
lines (PG~1114+445: Mathur, Wilkes \& Elvis 1998; 
MR~2251--178: Mathur et~al., in preparation), and it has been argued
that the X-ray and UV absorption arise in the same gas. While there is still
debate over the extent to which the X-ray and UV absorbers can be
unified, they are likely to have qualitatively similar dynamics. 
The UV absorbing gas is measured to be outflowing from the nucleus at 
speeds of several hundred km~s$^{-1}$. 

The incidence of warm absorbers in luminous, radio-quiet QSOs is difficult 
to address at present (e.g., George et~al. 1999). Warm 
absorbers are detected in $\simgt 50$\% of 
Seyfert~1s, while a much smaller percentage of radio-quiet QSOs have
{\it detected\/} warm absorbers. However, the X-ray spectra
of most radio-quiet QSOs have significantly lower signal-to-noise
than those of Seyfert~1s, and cosmological redshifting also moves the
main warm absorber edges down to regions of low effective area and
often poor calibration. Seyfert-like warm absorbers could be lurking
undetected in the noisy X-ray spectra of many radio-quiet QSOs, and
the only clear conclusion that can be drawn at present is that better
data are needed (although Laor et~al. 1997 suggest that warm absorbers
are relatively rare in radio-quiet QSOs based upon \rosat\ PSPC data).  

\section{Broad Absorption Line (BAL) QSOs}

Luminous radio-quiet QSOs show another type of absorption that is not
familiar from Seyfert~1s: UV Broad Absorption Lines (BALs) that are
created in an outflowing `wind' with velocities up to $\sim 0.1c$. 
BALs have been intensively studied in the UV for many years, and it is
likely that most QSOs create BAL outflows (e.g., Weymann 1997). The 
BAL region is thought to be major part of the nuclear environment
with a covering factor of 10--50\% 
(e.g., Goodrich 1997; Krolik \& Voit 1998), and the BAL phenomenon
may be fundamentally connected to the QSO `radio volume control' 
(e.g., Weymann 1997). In addition, BAL outflows may clear gas from
QSO host galaxies and thereby affect star formation and QSO fueling
over long timescales (e.g., Fabian 1999). 

Ideally, one would like to use X-rays to study the absorption properties, 
nuclear geometries, and continuum shapes of BAL~QSOs. X-ray absorption 
studies would constrain the column density, ionization state, abundances,
and covering factor of the BAL gas, and the nuclear geometry could be
constrained using the iron~K$\alpha$ line and X-ray variability. 
Regarding the continuum shape, it is important to establish that,
underneath their absorption, BAL~QSOs emit like normal radio-quiet 
QSOs. 

\rosat\ observations found BAL~QSOs to be very weak in the soft
X-ray band with few X-ray detections (e.g., Kopko, Turnshek \& Espey 1994; 
Green \& Mathur 1996, hereafter GM96). This was an important 
and surprising result since, if BAL~QSOs indeed have normal 
underlying X-ray continua, large neutral column densities of 
$\simgt 4\times 10^{22}$~cm$^{-2}$ are required to extinguish the X-ray 
emission. Ionization of the absorbing gas led to even larger required
column densities. The \rosat\ column densities were at least an order 
of magnitude larger than those inferred from UV data.
Subsequently, it was realized that the UV absorption
lines are severely saturated (e.g., Hamann 1998), leading to much larger
inferred UV column densities. These column densities are then consistent 
with the X-ray data.

\begin{figure}[t!]
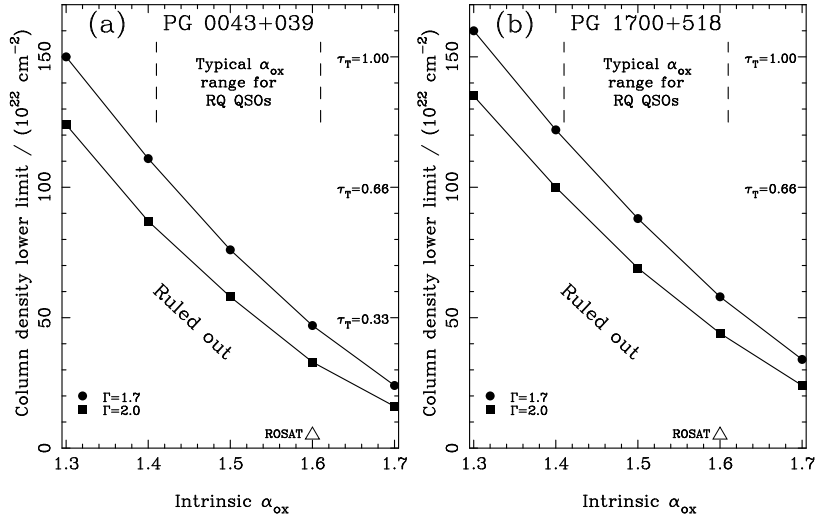

\centerline{\psfig{file=fig2a.ps, width=5.3cm}
\psfig{file=fig2b.ps, width=5.3cm}}
\caption[]{Column density lower limits for (a) PG~0043+039 and 
(b) PG~1700+518 derived using our \asca\ SIS0 data. We 
show the inferred column density lower limit as 
a function of the intrinsic (i.e., absorption-corrected) value of
\aox\ (the slope of a nominal power law connecting the rest-frame
flux density at 2500~\AA\ to that at 2~keV). The square data points
are for an X-ray photon index of $\Gamma=2.0$, and the circular
dots are for an X-ray photon index of $\Gamma=1.7$. The open triangle
at $\alpha_{\rm ox}=1.6$ illustrates the typical BAL~QSO column 
density lower limit found by GM96 based on \rosat\ data. The numbers
along the right-hand sides of the panels show the Thomson optical depth
of the corresponding column density; note that our inferred column
densities are within a factor of $\approx 3$ of being optically
thick to Thomson scattering. The column density lower limits 
shown in this plot are for absorption by neutral gas with solar
abundances. In reality, the absorbing gas is 
probably ionized, and this can significantly 
raise the required column density. We have made similar plots 
to those above using $\alpha_{\rm ix}$ (between 1.69~$\mu$m and 
2~keV), and we find similarly large column densities are required. 
From Gallagher et~al. (1999).}
\end{figure}

Hard X-rays are much more penetrating than soft X-rays, and if BAL~QSOs
are absorbed by column densities of a few times $10^{22}$~cm$^{-2}$ they 
should be much brighter above 2~keV than below this energy. 
Using \asca, Mathur, Elvis \& Singh (1995) detected the famous BAL~QSO
PHL~5200 ($z=1.98$) and claimed to measure a large column density for 
this object via spectral fitting. Our independent analysis of these data
confirms the detection, but the claim for a large column density 
in this object is not reliable at present due to extremely limited photon 
statistics (Gallagher et~al. 1999).
The nearby BAL~QSO Mrk~231 ($z=0.042$) has also been studied by \asca\
(Iwasawa 1999; Turner 1999) and appears to show absorption with a 
column density of $\simgt 2\times 10^{22}$~cm$^{-2}$, although precise
constraints are difficult due to the complex X-ray spectrum 
of this object (e.g., there appears to be a significant starburst 
contribution in X-rays). 

Recently, we have been performing an exploratory BAL~QSO survey using
\asca\ and \bepposax\ (Gallagher et~al. 1999; Gallagher et~al., 
in preparation). We chose these satellites because they provide access to 
penetrating 2--10~keV X-rays. We performed moderate-length 
($\approx$~20--30~ks) exploratory observations to learn about the basic 
X-ray properties of as many BAL~QSOs as possible without being too heavily 
invested in the uncertain results from any one object. Our goals were to 
define the 2--10~keV properties (e.g., fluxes) of the class, to discover 
good objects for follow-up studies with \chandra\ and \xmm, and to set 
absorption, geometry and continuum constraints (to the greatest extent 
possible with exploratory observations). We proposed many of the
optically brightest BAL~QSOs known since the optical and X-ray fluxes 
are generally correlated for QSOs. Most of our objects should have 
been easily detected if they have normal QSO X-ray continua absorbed by
column densities of several times $10^{22}$~cm$^{-2}$. We focused on
bona-fide BAL~QSOs (no mini-BALs; see \S3.1 of Weymann et~al. 1991), 
and we also tried to sample a few objects with extreme properties 
(e.g., optical continuum polarization) to look for correlations. 

We have performed new \asca\ and \bepposax\ observations for 8 BAL~QSOs in
total, and we have also analyzed the archival data for 4 BAL~QSOs.
Our objects have $z=$~0.042--3.505 and $B=$~14.5--18.5; 
PHL~5200 ($B=18.5$) is the optically faintest member of our sample. 
We detect 5 of our 12 BAL~QSOs, with our most distant and most
luminous detected BAL~QSO being CSO~755 ($z=2.88$, $M_{\rm V}=-27.4$; 
Brandt et~al. 1999). Our detection fraction is higher
than in soft X-rays, consistent with the idea that heavy absorption 
is present in these objects. However, we find that
BAL~QSOs are still generally faint 2--10~keV sources, and several 
of them are strikingly faint. For example, we did not detect the optically
bright BAL~QSOs PG~0043+039 ($B=15.9$, $z=0.384$, 24~ks \asca\ exposure) 
and PG~1700+518 ($B=15.4$, $z=0.292$, 21~ks \asca\ exposure). If these
objects have normal underlying X-ray continua, then large neutral column 
densities of $\simgt 5\times 10^{23}$~cm$^{-2}$ are needed to explain
their X-ray non-detections (see Figure~2). Because of our access to 
more penetrating X-rays, our column density lower limits for some 
objects are about an order of magnitude larger than those 
set by \rosat. Ionization of the absorbing
gas raises our required column densities to the point where they are
almost `Compton-thick' ($N_{\rm H}\simgt 1.5\times 10^{24}$~cm$^{-2}$;
compare with Murray et~al. 1995). These large column densities 
increase the inferred mass outflow rate and kinetic 
luminosity. If the X-ray absorption arises in gas 
at $\simgt 3\times 10^{16}$~cm that is outflowing with a significant 
fraction of the terminal velocity measured
from the UV BALs, one derives extremely large mass outflow rates
($\dot M_{\rm outflow}\simgt 5$~M$_\odot$~yr$^{-1}$) and 
kinetic luminosities ($L_{\rm kinetic}\simgt L_{\rm ionizing}$). 
While such powerful winds are perhaps not impossible, the 
mass outflow rate and kinetic luminosity
can be reduced if much of the X-ray absorption occurs at 
velocities significantly smaller than the BAL terminal velocity. 
Note that the X-ray and UV absorbers in BAL~QSOs have not yet been
shown to be identical, and the X-ray and UV light paths may differ.

Our exploratory observations demonstrate that it is risky to attempt 
long X-ray spectroscopic observations of BAL~QSOs that do not have 
established X-ray fluxes, and we find that optical flux is {\it not\/} a 
good predictor of X-ray flux for BAL~QSOs. We fail to detect some of
our optically brightest objects, while some of our optically faintest
are clearly detected. We have empirically searched for other predictors 
of X-ray brightness, and while the data are limited there is a
tentative connection between high optical continuum polarization 
and X-ray brightness (see Brandt et~al. 1999 for details). Such a 
connection could be physically understood if the direct lines of sight
into the X-ray nuclei of BAL~QSOs are usually blocked by 
Compton-thick matter, and we can only see X-rays when there is 
substantial electron scattering in the nuclear environment by a
`mirror' of moderate Thomson depth. Further studies of uniform,
well-defined BAL~QSO samples are needed to avoid biases and check 
this potential connection better. It can also be checked with
detailed X-ray studies of highly polarized BAL~QSOs. Iron~K$\alpha$ 
lines with large equivalent widths could be formed if most of the
X-ray flux is scattered, and one would also not expect rapid
($\simlt 1$~day) X-ray variability. 

\section{Soft X-ray Weak (SXW) QSOs}

BAL~QSOs are generally weak in the soft X-ray band, presumably due to
heavy X-ray absorption. One can also address the converse questions: 
Do all Soft X-ray Weak QSOs (SXW~QSOs) suffer from absorption? 
Do all SXW~QSOs have BALs? 
Alternative possible causes of soft X-ray weakness include unusual 
intrinsic spectral energy distributions (SEDs) and extreme X-ray or 
optical variability (e.g., changes in \aox\ over time). 
The presence of QSOs with relatively weak soft X-ray emission was
recognized at least as early as the mid-1980s, with 
some observed to be $\simgt 20$ times weaker than expected given 
their optical fluxes (e.g., Elvis \& Fabbiano 1984; 
Avni \& Tananbaum 1986; Elvis 1992). 
For example, Avni \& Tananbaum (1986) discussed a `skew tail' towards
soft X-ray weak objects for the \aox\ distribution of the PG QSOs. 
Many new SXW~QSOs were found in \rosat\ samples 
(e.g., Laor et~al. 1997; Yuan et~al. 1998), and \rosat\ was also
able to place significantly tighter constraints upon \aox. 
This sparked further detailed studies of these objects
(e.g., Wang et~al. 1999; Wills, Brandt \& Laor 1999), and 
we have recently completed the first systematic study of a well-defined
SXW~QSO sample (Brandt, Laor \& Wills 1999). Our goals for this
study were 
(1) to determine the origin of soft X-ray weakness in general,
(2) to discover relations between SXW~QSOs, BAL~QSOs, 
and X-ray warm absorber QSOs, and
(3) to search for correlations between soft X-ray weakness and other
interesting observables. 

We selected all SXW~QSOs from the Boroson \& Green (1992, hereafter BG92) 
sample of 87 $z<0.5$ PG QSOs. The BG92 sample is well defined and 
representative of the optically selected QSO population, and there 
is already a large amount of high-quality and uniform data available 
for it. 
We computed our own \aox\ values for the BG92 objects using data mainly 
from \rosat\ but also from \asca\ and \einstein\ as needed, and our
resulting \aox\ values were substantially more complete and 
constraining than those previously available (especially for the
SXW~QSOs). 
We used $\alpha_{\rm ox}\leq-2$ as our criterion for soft X-ray 
weakness (note in this section we take \aox\ to be a {\it negative\/}
quantity). Thus, given their optical fluxes, our SXW~QSOs were $\geq 25$ 
times weaker than `usual' in soft X-rays.
We found 10 SXW~QSOs with $\alpha_{\rm ox}\leq-2$, and thus 
SXW~QSOs appear to comprise $\approx 11$\% of the optically selected
QSO population. Nine of our SXW~QSOs are radio-quiet, and one is
radio-loud. 

\begin{figure}[t!]
\centerline{\psfig{file=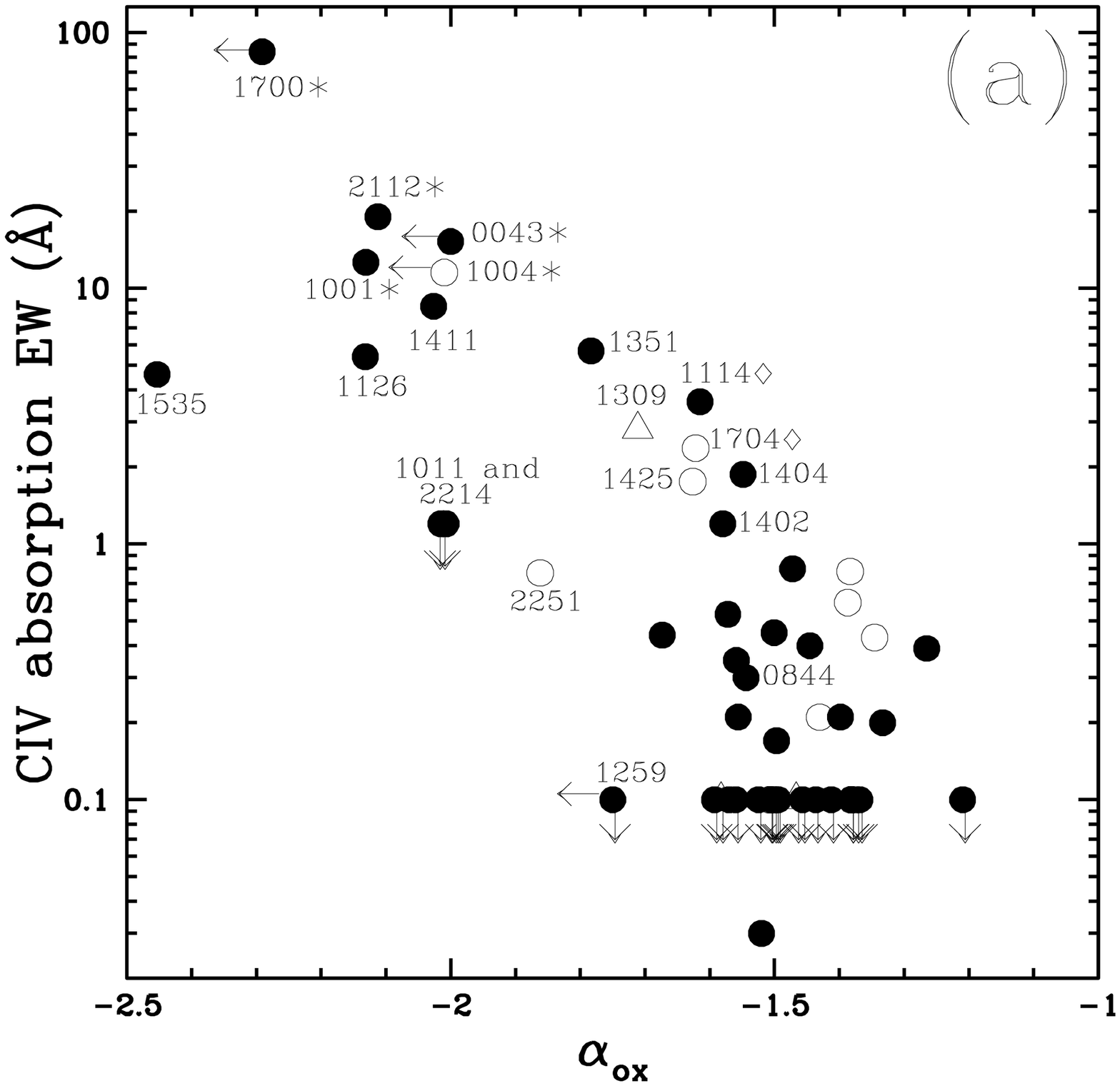, width=7cm}
\psfig{file=fig3b.ps, width=5.5cm}}
\caption[]{(a) Plot of \aox\ versus 
C~{\sc iv}~$\lambda 1549$ 
absorption-line EW (in the
rest frame) for the BG92 QSOs with UV coverage. Following BG92, 
solid dots are radio-quiet QSOs, 
open triangles are core-dominated radio-loud QSOs, and
open circles are lobe-dominated radio-loud QSOs. 
SXW~QSOs (those QSOs with $\alpha_{\rm ox}\leq -2$)
and other interesting QSOs are labeled by the 
right ascension part of their name. An asterisk to the right
of an object's name indicates that it is a BAL~QSO or probable 
BAL~QSO, and a diamond indicates that it is known to have an 
X-ray warm absorber. 
(b) UV spectra of the regions of the broad C~{\sc iv} emission
line for some of the SXW~QSOs. Note the strong, blueshifted
C~{\sc iv} absorption. From Brandt, Laor \& Wills (1999).}
\end{figure}

We compared the continuum and line properties of our 10 SXW~QSOs to
those of the other 77 BG92 non-SXW~QSOs using nonparametric tests.
The properties compared included those listed in 
Tables~1 and 2 of BG92 as well as the optical continuum polarization,
the optical continuum slope, and the radio structure. We also
compared the C~{\sc iv}~$\lambda 1549$ 
absorption-line properties for the 55 QSOs
from BG92 that have high-quality UV coverage in this spectral
region. All C~{\sc iv} measurements were made by B.\,J. Wills with
particular effort toward ensuring consistency and uniformity.
We found that the SXW~QSOs and non-SXW~QSOs have consistent
distributions of $M_{\rm V}$, $z$, radio loudness ($R$), 
optical continuum slope,
optical continuum polarization, and H$\beta$ FWHM. In addition,
they have consistent EW distributions of H$\beta$, He~{\sc ii} and
Fe~{\sc ii}. SXW~QSOs were found to have significantly lower
[O~{\sc iii}] luminosities than those of non-SXW~QSOs; low
[O~{\sc iii}] luminosities have similarly been noted for 
low-ionization BAL~QSOs (e.g., Turnshek et~al. 1997). 
Since [O~{\sc iii}] emission is likely to be a reasonably 
isotropic property for radio-quiet PG QSOs (e.g., Kuraszkiewicz et~al. 1999),
this result is significant as it suggests there may be an {\it intrinsic\/}
difference between SXW~QSOs and non-SXW~QSOs (see Brandt, Laor \& Wills 1999). 
In addition, SXW~QSOs appear to have `peaky' H$\beta$ line profiles and 
large H$\beta$ line shifts (either to the blue or the red relative to
the rest frame defined by [O~{\sc iii}]). 

The most striking difference between the SXW~QSOs and 
non-SXW~QSOs is their UV absorption. SXW~QSOs show
greatly enhanced C~{\sc iv} absorption 
(see Figure~3). We find blueshifted C~{\sc iv} absorption with 
EW~$>4.5$~\AA\ in 8 of our 10 SXW~QSOs, while only 1 of 45 
non-SXW~QSOs had EW~$>4.5$~\AA. The two SXW~QSOs without clear 
UV absorption, 1011--040 and 2214+139, have UV spectra of only 
limited quality. Given that UV and X-ray absorption 
have a high probability of joint occurrence in Seyfert galaxies 
and QSOs, we consider Figure~3 to be evidence 
that absorption is the primary cause of soft X-ray weakness 
in QSOs. Only one of our SXW~QSOs, 1411+442, has a broad-band
X-ray spectrum at present, and it indeed shows evidence for strong
X-ray absorption with $N_{\rm H}\simgt 10^{23}$~cm$^{-2}$
(Brinkmann et~al. 1999).
We can argue against unusual SEDs as the primary
cause of soft X-ray weakness by noting that our SXW~QSOs have 
normal H$\beta$, He~{\sc ii} and Fe~{\sc ii} EWs (see 
Korista, Ferland \& Baldwin 1997). We also do not find general
evidence for strong \aox\ variability when we compare our
\aox\ values with the limited historical data available.  
The fact that we find no evidence for QSOs with intrinsically 
weak soft X-ray emission underscores the universality of QSO X-ray 
production. 

The general correlation between \aox\ and C~{\sc iv} absorption EW 
shown in Figure~1 provides a useful overall view of QSO absorption. 
Unabsorbed QSOs and BAL~QSOs lie at opposite extremes of the 
correlation, while X-ray warm absorber QSOs and moderate SXW~QSOs
lie at intermediate positions. We find 4--5 bona-fide BAL~QSOs in 
the BG92 sample; 3 were already known (0043+039, 1700+518 and 2112+059) 
and 2 are new (1001+054 and probably 1004+130; see 
Wills, Brandt \& Laor 1999 for detailed discussion of 
1004+130). If all BAL~QSOs 
are SXW~QSOs, then we should have found all the BAL~QSOs in the 
BG92 sample. The incidence of BAL~QSOs in the BG92 sample appears 
to be statistically consistent with the $\approx 11$\% 
observed for the LBQS (e.g., Weymann 1997), although this issue 
could be examined more reliably with complete UV coverage of the
BG92 QSOs. 

The UV results for our SXW~QSOs 
imply that selection by soft X-ray weakness is an effective 
($\sim 80$\% successful) way to find low-redshift QSOs with strong 
UV absorption. This is important from a practical 
point of view because, for bright QSOs, the optical and X-ray flux 
densities needed to establish soft X-ray weakness can often be obtained 
from publicly available data. This method 
has already been exploited in several cases for individual objects 
(e.g., Fiore et~al. 1993; Mathur et~al. 1994), and it could be 
profitably applied to larger QSO samples. 

\section{Some Future Prospects}

With the next generation of X-ray observatories, it should be
possible to find many more X-ray warm absorbers in radio-quiet
QSOs or demonstrate that few are present. This will allow
study of their basic physical properties as well as a reliable
determination of their incidence. Detailed X-ray spectroscopy
and modeling should be possible for a few of the X-ray brightest
sources, although this will require a significant investment of
observation time.  

For BAL~QSOs, further exploratory observations are needed to 
look for correlations with optical continuum polarization and
other properties. These would be most effective if performed on
uniform and well-defined samples. Moderate-quality X-ray spectroscopy 
should be possible for a few of the X-ray brightest BAL~QSOs to study
their absorption properties, nuclear geometries, and  
continuum shapes. For BAL~QSOs with enough X-ray flux, 
the widths and amplitudes of X-ray bound-free edges can constrain 
the dynamics and metallicity of the absorber. It is 
also important to study the radio-loud
BAL~QSOs (e.g., Becker et~al. 1999) in X-rays to determine if 
they follow the same patterns as radio-quiet objects, and deep 
X-ray surveys over moderate areas may be able to constrain 
the BAL covering factor (see Krolik \& Voit 1998). 
Studies of SXW~QSOs more generally would benefit from a focused
X-ray and UV study of a complete sample. The PG SXW~QSOs are
probably a good starting point, but a larger complete sample 
would be even better. Intense studies of particularly interesting
objects (e.g., 1004+130) are important as well.

Finally, it is crucial to test models that 
propose to unify the different types of X-ray absorption
into a coherent physical picture. Such testing should provide an
exciting challenge for even the next generation of X-ray 
observatories. 

\begin{acknowledgements}
We acknowledge the support of 
NASA LTSA grant NAG5-8107 and the Alfred P. Sloan Foundation (WNB),
NASA grant NAG5-4826 and the Pennsylvania Space Grant Consortium (SCG), 
the fund for the promotion of research at the Technion (AL), and
NASA LTSA grant NAG5-3431 (BJW). We thank C.S. Reynolds for a careful
reading and J. Chiang for a helpful discussion.  
\end{acknowledgements}


\begin{references}

\par\noindent\hangindent 15pt Avni, Y., Tananbaum, H. 1986, ApJ, 305, 83

\par\noindent\hangindent 15pt Becker, R.H., et~al. 1999, ApJ, in preparation

\par\noindent\hangindent 15pt Boroson, T.A., Green, R.F. 1992, ApJS, 80, 109 (BG92)

\par\noindent\hangindent 15pt Brandt, W.N., Fabian, A.C., Pounds, K.A. 1996, MNRAS, 278, 326

\par\noindent\hangindent 15pt Brandt, W.N., Mathur, S., Reynolds, C.S., Elvis, M. 1997, MNRAS, 292, 407

\par\noindent\hangindent 15pt Brandt, W.N., Laor, A., Wills, B.J. 1999, ApJ, in press (astro-ph/9908016)

\par\noindent\hangindent 15pt Brandt, W.N., Comastri, A., Gallagher, S.C., Sambruna, R.M., Boller, Th., 
Laor, A. 1999, ApJ, in press (astro-ph/9909284)

\par\noindent\hangindent 15pt Brinkmann, W., Wang, T., Matsuoka, M., Yuan, W. 1999, A\&A, 345, 43

\par\noindent\hangindent 15pt Crenshaw, D.M., Kraemer, S.B., Boggess, A., Maran, S.P., Mushotzky, R.F., Wu, C. 
1999, ApJ, 516, 750

\par\noindent\hangindent 15pt Elvis, M., Fabbiano, G. 1984, ApJ, 280, 91

\par\noindent\hangindent 15pt Elvis, M. 1992, 
in Frontiers of X-ray Astronomy,
ed. Tanaka, Y., Koyama, K. 
(Universal Acad. Press, Tokyo), p. 567

\par\noindent\hangindent 15pt Fabian, A.C. 1999, MNRAS, in press (astro-ph/9908064)

\par\noindent\hangindent 15pt Fiore, F., Elvis, M., Mathur, S., Wilkes, B.J., McDowell, J.C.
1993, ApJ, 415, 129

\par\noindent\hangindent 15pt Gallagher, S.C., Brandt, W.N., Sambruna, R.M., Mathur, S., Yamasaki, N.
1999, ApJ, 519, 549 

\par\noindent\hangindent 15pt George, I.M., et~al. 1997, ApJ, 491, 508

\par\noindent\hangindent 15pt George I.M., Turner T.J., Netzer H., Nandra K., Mushotzky R.F., Yaqoob T. 
1998, ApJS, 114, 73

\par\noindent\hangindent 15pt George, I.M., et~al. 1999, ApJ, in press (astro-ph/9910218)

\par\noindent\hangindent 15pt Goodrich, R.W. 1997, ApJ, 474, 606

\par\noindent\hangindent 15pt Green, P.J., Mathur, S. 1996, ApJ, 462, 637 (GM96)

\par\noindent\hangindent 15pt Grupe, D., Wills, B.J., Wills, D., Beuermann, K. 1998, A\&A, 333, 827

\par\noindent\hangindent 15pt Halpern, J.P. 1984, ApJ, 281, 90

\par\noindent\hangindent 15pt Hamann, F. 1998, ApJ, 500, 798

\par\noindent\hangindent 15pt Iwasawa, K. 1999, MNRAS, 302, 96

\par\noindent\hangindent 15pt Kopko, M., Turnshek, D.A., Espey, B.R. 1994, 
in Multi-Wavelength Continuum Emission of AGN,
ed. Courvoisier, T., Blecha, A. 
(Kluwer, Dordrecht), p. 450

\par\noindent\hangindent 15pt Korista, K., Ferland, G., Baldwin, J. 1997, ApJ, 487, 555

\par\noindent\hangindent 15pt Krolik, J.H., Voit, G.M. 1998, ApJ, 497, L5

\par\noindent\hangindent 15pt Kuraszkiewicz, J., Wilkes, B.J., Brandt, W.N., Vestergaard, M. 1999, ApJ, submitted

\par\noindent\hangindent 15pt Laor, A., Fiore, F., Elvis, M., Wilkes, B.J., McDowell, J.C. 1994, ApJ, 435, 611 

\par\noindent\hangindent 15pt Laor, A., Fiore, F., Elvis, M., Wilkes, B.J., McDowell, J.C. 1997, ApJ, 477, 93 

\par\noindent\hangindent 15pt Leighly, K.M., Kay, L.E., Wills, B.J., Wills, D., Grupe, D. 1997, ApJ, 489, L25 

\par\noindent\hangindent 15pt Mathur, S., Wilkes, B.J., Elvis, M., Fiore, F. 1994, ApJ, 434, 493

\par\noindent\hangindent 15pt Mathur, S., Elvis, M., Singh, K.P. 1995, ApJ, 455, L9

\par\noindent\hangindent 15pt Mathur, S., Wilkes, B.J., Elvis, M. 1998, ApJ, 503, L23 

\par\noindent\hangindent 15pt Murray, N., Chiang, J., Grossman, S.A., Voit, G.M. 1995, ApJ, 451, 498

\par\noindent\hangindent 15pt Pan, H.C., Stewart, G.C., Pounds, K.A. 1990, MNRAS, 242, 177

\par\noindent\hangindent 15pt Reeves, J.N., Turner, M.J.L., Ohashi, T., Kii, T. 1997, MNRAS, 292, 468

\par\noindent\hangindent 15pt Reynolds, C.S. 1997, MNRAS, 286, 513

\par\noindent\hangindent 15pt Siebert, J., Komossa, S., Brinkmann, W. 1999, A\&A, in press 
(astro-ph/9909323)

\par\noindent\hangindent 15pt Turner, T.J. 1999, ApJ, 511, 142 

\par\noindent\hangindent 15pt Turnshek, D.A., Monier, E.M., Sirola, C.J., Espey, B.R. 1997, ApJ, 476, 40

\par\noindent\hangindent 15pt Wang, T.G., Brinkmann, W., Wamsteker, W., Yuan, W., Wang, J.X. 1999, MNRAS, 307, 821 

\par\noindent\hangindent 15pt Weymann, R.J., Morris, S.L., Foltz, C.B., Hewett, P.C. 1991, ApJ, 373, 23

\par\noindent\hangindent 15pt Weymann, R.J. 1997, 
in Mass Ejection from AGN, 
ed. Arav, N., Shlosman, I., Weymann, R.J. 
(ASP Press: San Francisco), p. 3

\par\noindent\hangindent 15pt Wills, B.J., Brandt, W.N., Laor, A. 1999, ApJ, 520, L91 

\par\noindent\hangindent 15pt Yuan, W., Brinkmann, W., Siebert, J., Voges, W. 1998, A\&A, 330, 108

\end{references}
\end{document}